\documentstyle[12pt]{article}
\input html.sty
\begin{document}
\title{MATTERS OF GRAVITY, The newsletter of the APS Topical Group 
on Gravitation}
\begin{center}
{ \Large {\bf MATTERS OF GRAVITY}}\\
\bigskip
\hrule
\medskip
{The newsletter of the Topical Group on Gravitation of the American Physical 
Society}\\
\medskip
{\bf Number 13 \hfill Spring 1999}
\end{center}
\begin{flushleft}

\tableofcontents
\bigskip
\hrule
\vfill

\section*{\noindent  Editor\hfill}

\medskip
Jorge Pullin\\
\smallskip
Center for Gravitational Physics and Geometry\\
The Pennsylvania State University\\
University Park, PA 16802-6300\\
Fax: (814)863-9608\\
Phone (814)863-9597\\
Internet: 
\htmladdnormallink{\protect {\tt pullin@phys.psu.edu}}
{mailto:pullin@phys.psu.edu}\\
WWW: \htmladdnormallink{\protect {\tt http://www.phys.psu.edu/\~{}pullin}}
{http://www.phys.psu.edu/~pullin}
\begin{rawhtml}
<P>
<BR><HR><P>
\end{rawhtml}
\end{flushleft}
\vfill
\pagebreak

\section*{Editorial}

I just wanted to renew the invitation to everyone to suggest articles
for the newsletter. The only way to keep the newsletter vibrant and
balanced is if I hear from you, don't hesitate to email me with 
suggestions.

The next newsletter is due February 1st.  If everything goes well this
newsletter should be available in the gr-qc Los Alamos archives under
number gr-qc/9902011. To retrieve it send email to 
\htmladdnormallink{gr-qc@xxx.lanl.gov}{mailto:gr-qc@xxx.lanl.gov}
(or 
\htmladdnormallink{gr-qc@babbage.sissa.it}{mailto:gr-qc@babbage.sissa.it} 
in Europe) with Subject: get 9902011
(numbers 2-10 are also available in gr-qc). All issues are available in the
WWW:\\\htmladdnormallink{\protect {\tt
http://vishnu.nirvana.phys.psu.edu/mog.html}}
{http://vishnu.nirvana.phys.psu.edu/mog.html}\\ 
A hardcopy of the newsletter is
distributed free of charge to some members of the APS
Topical Group on Gravitation. It is considered a lack of etiquette to
ask me to mail you hard copies of the newsletter unless you have
exhausted all your resources to get your copy otherwise.

If you have comments/questions/complaints about the newsletter email
me. Have fun.
\bigbreak

\hfill Jorge Pullin\vspace{-0.8cm}
\vfill
\pagebreak
\section*{\noindent Editorial policy:}

The newsletter publishes articles in three broad categories, 

1. News about the topical group, normally contributed by officers of the 
group.

2. Research briefs, comments about new developments in research,
typically by an impartial observer. These articles are normally
by invitation, but suggestions for potential topics and authors are welcome
by the correspondents and the editor.

3. Conference reports, organizers are welcome to contact the editor or
correspondents, the reports are sometimes written by participants in the
conference in consultation with organizers.

Articles are expected to be less than two pages in length in all categories.

Matters of Gravity is not a peer-reviewed journal for the publication
of original research. We also do not publish full conference or meeting
announcements, although  we might consider publishing a brief notice 
with indication of a web page or other contact information.

\section*{Correspondents}
\begin{itemize}
\item John Friedman and Kip Thorne: Relativistic Astrophysics,
\item Raymond Laflamme: Quantum Cosmology and Related Topics
\item Gary Horowitz: Interface with Mathematical High Energy Physics and
String Theory
\item Richard Isaacson: News from NSF
\item Richard Matzner: Numerical Relativity
\item Abhay Ashtekar and Ted Newman: Mathematical Relativity
\item Bernie Schutz: News From Europe
\item Lee Smolin: Quantum Gravity
\item Cliff Will: Confrontation of Theory with Experiment
\item Peter Bender: Space Experiments
\item Riley Newman: Laboratory Experiments
\item Warren Johnson: Resonant Mass Gravitational Wave Detectors
\item Stan Whitcomb: LIGO Project
\item Peter Saulson; former editor, correspondent at large.
\end{itemize}
\vfill
\pagebreak


\section*{\centerline {Topical group news}}
\addtocontents{toc}{\protect\smallskip}
\addtocontents{toc}{\bf News:}
\addtocontents{toc}{\protect\smallskip}
\addcontentsline{toc}{subsubsection}{\it  Topical group news, by Jim Isenberg}
\begin{center}
    Jim Isenberg, TGG secretary, University of Oregon\\
\htmladdnormallink{jim@newton.uoregon.edu}
{mailto:jim@newton.uoregon.edu}
\end{center}
\parindent=0pt
\parskip=5pt

$\bullet$ Election News

The ballot for this year's election is complete, and it will very soon be
going out to all members. The list of candidates is as follows:

Vice Chair:
  Richard Matzner, 
  Bob  Wald.

Secretary/Treasurer:
  David Garfinkle,
  Ted Jacobson.

Delegate (Slot 1)
  John Friedman,
  Jennie Traschen.

Delegate (Slot 2)
  Matt Choptuik,
  Ed Seidel.

PLEASE VOTE.

$\bullet$ Annual Meeting

This year our annual meeting will be earlier than usual. It will be in
conjunction with the APS Centennial Meeting, 20-26 March, 1999, in Atlanta,
Georgia. This meeting is going to be very big. There will be a lot of
bicentennial stuff (exhibits, lectures, historical talks), and it should
be quite nice. Please check the web site:
\htmladdnormallink{http://www.aps.org/centennial}
{http://www.aps.org/centennial}

The activities tied in with GR and gravitation include the following

\bigbreak

{\it Centennial Symposium:  Einstein's Legacy: Nature's Experiments in
Gravitational Physics}   (Tuesday, 24 March)

Clifford Will:
 Einstein's Relativity Put to Nature's Test: A Centennial Perspective

Robert Kirshner:
 Was the Cosmological Constant Einstein's Greatest Mistake?

David Spergel:
  The Cosmic Microwave Background: A Bridge to the Early Universe

Kip Thorne:
 A New Window on the Universe: The Search for Gravitational Waves

\bigbreak

{\it Invited Session:   Progress in the theory of gravitation}  
(Thursday, 26 March)

Robert Wald:
 Classical general relativity

Saul Teukolsky:
 Numerical methods

Gary Horowitz:
 Quantum gravity

\bigbreak

{\it Invited Session: Instrumentation for Gravitational Radiation Detection}
                                 ( Monday,  23 March)

David Shoemaker: Interferometric Detectors - overview

Peter Fritschel: Gravitational wave interferometer configurations

Jordan Camp: 
 Requirements and performance of optics for gravitational wave interferometry

Eric Gustafson:
  Lasers for Gravitational wave interferometry

Peter Saulson:
Thermal noise in gravitational wave interferometers

There will also be a number of focus sessions with invited and contributed
talks (See www site), plus the
Annual TGG Business Meeting, which will be on Wednesday at 5:30.
PLEASE COME.

$\bullet$  Travel Support for the Annual Meeting

Last year, we decided that a useful way to spend our accumulating treasure
is to support young people coming to the annual meeting. So this year, we
will begin a program of small grants to help pay travel and/or lodging for
the meeting in Atlanta.  The people eligible for these grants are students
and post docs.  If you wish to apply, please send the following to Jim
Isenberg (\htmladdnormallink{jim@newton.uoregon.edu}
{mailto:jim@newton.uoregon.edu}).

1) A brief CV

2) A list of your publications

3) A supporting letter from a faculty member research group director

4) A tentative travel plan (including rough cost) for going to the meeting

The applications should be in by the 20th of  February.

\vfill
\pagebreak

\section*{\centerline {We hear that...}}
\addcontentsline{toc}{subsubsection}{\it  We hear that..., by Jorge Pullin}
\begin{center}
    Jorge Pullin, Editor\\
\htmladdnormallink{pullin@psu.edu}
{mailto:pullin@psu.edu}
\end{center}
\parindent=0pt
\parskip=5pt

{\bf Stephen Hawking} has received the Lilienthal prize from APS ``For boldness
and creativity in gravitational physics, best illustrated by the
prediction that black holes should emit black body radiation and
evaporate, and for the special gift of making abstract ideas accessible
and exciting to experts, generalists, and the public alike."

{\bf Luis Lehner} has received the Metropolis award from APS ``For developing a
method that significantly advances the capability for modeling
gravitational radiation by making possible the stable numerical solution
of Einstein's equation near moving black holes."

{\bf Beverly Berger} was elected fellow of the APS ``For her pioneering
contributions to global issues in classical general relativity,
particularly the analysis of the nature of cosmological singularities,
and for founding the Topical Group on Gravitation of the APS."

{\bf Joan Centrella} was elected fellow of the APS `` For her original
contributions to numerical relativity, cosmology, and astrophysics, in
particular for her studies of large-scale structure in the universe
and sources of gravitational radiation."

{\bf Ron Drever} was elected fellow of the APS ``For his fundamental experiment
to test the isotropy of space, and for his pioneering contributions to
laser interferometry as a tool for gravitational-wave detection."

{\bf Bernie Schutz} was elected fellow of the APS ``For his pioneering work in
the theory of gravitational radiation, for the discovery of new
instabilities in rotating, relativistic stars, and for elucidating how
gravitational-wave observations can reveal astrophysical and
cosmological information."

{\bf Stu Shapiro} was elected fellow of the APS ``For his broad contributions
to theoretical astrophysics and general relativity,
 including the physics of black holes, neutron stars, and large N-body
dynamical systems, and his pioneering use of supercomputers to explore
these areas. "

Our warmest congratulations to them all!

\vfill
\pagebreak

\section*{\centerline {The Chandra Satellite}}
\addcontentsline{toc}{subsubsection}{\it  The Chandra Satellite, by
Beverly Berger}
\begin{center}
    Beverly Berger, Oakland University\\
\htmladdnormallink{berger@oakland.edu}
{mailto:berger@oakland.edu}
\end{center}
\parindent=0pt
\parskip=3pt

Those of you who knew Chandra and those who only knew of
him will be pleased to learn that NASA has named its soon-to-be-launched
Advanced X-ray Astrophysics Facility (formerly AXAF) the Chandra X-ray
Observatory. The Chandra Observatory will join the Hubble Space Telescope
and the Compton Gamma-ray Observatory in NASA's program of major
space-based astronomical facilities. The final observatory in the
series,  for infrared astronomy (SIRTF), is under development.

The X-ray telescope in the
Observatory has a larger collecting area (400 cm$^2$ at 1 KeV) and
significantly better angular resolution ($.5''$) than  previous X-ray
telescopes such as those on the Einstein and Rosat observatories.
Instruments include a CCD Imaging Spectrometer, developed by Penn State
and MIT, and a High Resolution Camera, built by the Smithsonian
Astrophysical Observatory. After Space Shuttle deployment, rockets will
boost the Chandra Observatory into an unusual elliptical orbit with
apogee more than 1/3 the distance to the  moon. This will allow it to
spend most of its time above the earth's radiation belts.  The Chandra
X-ray Observatory Center (CXC), operated by SAO, will control science and
flight operations of the Observatory.
Excerpts from the NASA press release are given below. For more information
see the Chandra Observatory Web Site at 
\htmladdnormallink{http://xrtpub.harvard.edu/pub.html}
{http://xrtpub.harvard.edu/pub.html}.

{\it NASA's Advanced X-ray Astrophysics Facility has been renamed the
Chandra X-ray Observatory in honor of the late Indian-American Nobel
laureate, Subrahmanyan Chandrasekhar. The telescope is scheduled to be
launched no earlier than April 8, 1999 aboard the Space Shuttle Columbia
mission STS-93, commanded by astronaut Eileen Collins.

Chandrasekhar, known to the world as Chandra, which means ``moon" or
``luminous" in Sanskrit, was a popular entry in a recent NASA contest to
name the spacecraft. The contest drew more than six thousand entries from
fifty states and sixty-one countries. The co-winners were a tenth grade
student in Laclede, Idaho, and a high school teacher in Camarillo, CA.

``Chandra is a highly appropriate name," said Harvey Tananbaum, Director
of the CXC. ``Throughout his life Chandra worked tirelessly and with
great precision to further our understanding of the universe. These same
qualities characterize the many individuals who have devoted much of their
careers to building this premier x-ray observatory."

``Chandra probably thought longer and deeper about our universe than
anyone since Einstein," said Martin Rees, Great Britain's Astronomer
Royal.

``Chandrasekhar made fundamental contributions to the theory of black
holes and other phenomena that the Chandra X-ray Observatory will study.
His life and work exemplify the excellence that we can hope to achieve
with this great observatory," said NASA Administrator Dan Goldin.

Widely regarded as one of the foremost astrophysicists of the 20th century,
Chandrasekhar won the Nobel Prize in 1983 for his theoretical studies of
physical processes important to the structure and evolution of stars. He and
his wife immigrated from India to the U.S. in 1935. Chandrasekhar served
on the faculty of the University of Chicago until his death in 1995.

The Chandra X-ray Observatory will help astronomers worldwide better
understand the structure and evolution of the universe by studying
powerful sources of X rays such as exploding stars, matter falling into
black holes and other exotic celestial objects. X-radiation is an invisible
form of light produced by multi-million degree gas. Chandra will provide
x-ray images that are fifty times more detailed than previous missions. At
more than 45 feet in length and weighing more than five tons, it will be one
of the largest objects ever placed in Earth orbit by the Space Shuttle.}

\vfill
\pagebreak

\section*{\centerline {Analytical event horizons of merging black holes}}
\addtocontents{toc}{\protect\smallskip}
\addtocontents{toc}{\bf Research Briefs:}
\addtocontents{toc}{\protect\smallskip}
\addcontentsline{toc}{subsubsection}{\it  
Analytical event horizons of merging black holes, by Simonetta Frittelli}
\begin{center}
    Simonetta Frittelli, Duquesne University\\
\htmladdnormallink{simo@mayu.physics.duq.edu}
{mailto:simo@mayu.physics.duq.edu}
\end{center}
\parindent=0pt
\parskip=5pt

Probably all of us are familiar with the celebrated picture of the
pair of pants that made it to the cover of the November issue of {\it
Science} of 1995 [1].  This is the article where the
geometry of the collision of two black holes is explained in rather
lay terms (which has a definite appeal) and where an embedded picture
of the event horizon of such a collision was generated by numerically
integrating the light rays that generate the horizon.

But since I wish to make a point, it wouldn't be wise for me to leave
up to the reader to retrieve his own copy of the journal because, if
your desk is like mine, your copy must be buried beneath one of those
stacks of paper. Or, unfortunately enough, it could even have been
filed so ingeniously that you are sure now not to be able to find it.
Let me instead save you the trip to the library; to start with, let me
just borrow from the article the main features of event horizons:
({\it i\/}) The horizon is generated by null rays that continue
indefinitely into the future. ({\it ii\/}) The null generators may
either continue indefinitely into the past, or meet other generators
at points that are thereby considered as their starting points. ({\it
iii\/}) The cross-sectional area of the horizon increases
monotonically to a constant at late times.  We might as well say that
the event horizon is a null hypersurface that does not self-intersect,
which is formed by following each null geodesic in a bundle of finite
area into the past just so far up to the point where it meets another
geodesic of the bundle.  In fact, if we have an expanding null
hypersurface of finite area at late times, which generically does
self-intersect in the past, we might as well regard as an event
horizon the piece of the null hypersurface that lies to the future of
the crossovers, and regard the crossovers as the boundary of the
horizon.

In this context the ``seam'' along the inside of the trouser legs is a
crossover line where the generators are terminated.  The computer
simulation [2] of the horizon provided deep insight into the nature of
this boundary of the event horizon, distinguishing the caustic points
(where neighboring rays meet) from the simple crossover points (where
distant rays intersect without focusing).

The news is that {\it another\/} pair of pants has recently been
released.  It looks pretty much like the original, up to smooth
deformations.  My point is, however, that this newest pair of pants is
not the product of numerical integration, but is the embedded picture
of an analytical event horizon.  There is now an analytical expression
for the intrinsic metric of the event horizon of merging eternal black
holes.

The new pair of pants was constructed by Luis Lehner, Nigel Bishop,
Roberto G\'{o}mez, B\'{e}la Szil\'{a}gyi and Jeff Winicour [3], of the
University of Pittsburgh relativity group, which has traditionally
sustained an interest in null hypersurfaces (tell me about it).  The
recipe for making this event horizon calls for all sorts of
ingredients available in the pantry of the characteristic formulation
of the Einstein equations. Surprisingly, perhaps, it does not call for
a spacetime metric. Surprisingly, because one might think that, since
the metric is needed in order to find geodesics, the horizon could
only be known {\it a posteriory\/} of finding the spacetime metric.

The key to this remarkable work is to understand that the event
horizon can be used as partial data for constructing the spacetime
metric.  From this point of view, the metric will be known {\it a
posteriori\/} of finding the horizon!  And the horizon is found by
solving only constraint equations, namely, equations interior to the
horizon itself.

More precisely, the horizon is regarded as one of two intersecting
null hypersurfaces that jointly act as the initial surface for
evolution in double null coordinates.  In this case, the conformal
metric of the null slice constitutes free data.  The authors choose
the conformal structure so that the 3-metric of the horizon is
$\gamma_{ij}=\Omega^2h_{ij}$ where $h_{ij}$ is the pullback of the
Minkowski metric to a self-intersecting hypersurface which is null
with respect to the Minkowski metric. (An example of such a
self-intersecting hypersurface is the hypersurface traced in four
dimensions by the imploding wavefront of an ellipsoid in 3-space.)
The conformal factor is then determined by the projection
$n^an^bR_{ab}=0$ of the vacuum Einstein equations along the null
generators $n^a$ of the hypersurface.  This is an {\it ordinary\/}
second-order differential equation for $\Omega$ that determines the
dependence of $\Omega$ on the affine parameter $u$ along one null
geodesic.  Apparently, finding the solution is quite simple. The
freedom is huge, but the authors point out that $\Omega^2$ relates to
the cross-sectional area of the light beam, and thus its asymptotic
behavior is fixed by the condition that the area must be finite at
late times $u\to\infty$.  Furthermore, the behavior of the area
element at the boundary of the horizon is determined by the property
of the boundary of containing either caustic points or plain
crossovers, which is also used in restricting the behavior of
$\Omega^2$.  The requirement that the Weyl curvature must be regular
provides further tips for the integration.  The intrinsic geometry
$\gamma_{ij}$ of the horizon is thus found explicitly in terms of two
angular coordinates $(\theta,\phi)$ labeling the light rays, and the
affine parameter $u$, acting as a time.

It is rather instructive to see how the figure arises.  The pair of
pants is constructed by stacking up 3-dimensional Euclidean embeddings
of 2-dimensional surfaces obtained by slicing the horizon with
constant-$u$ hypersurfaces.  Actually, the figure corresponds to a
case of symmetry of revolution, so that one dimension can be ignored,
but this is exactly as in the case of the ``computational'' pair of
trousers of the {\it Science\/} article.  Also, strictly speaking, the
calculation represents the fission of two white holes, but time
reversion allows for its interpretation in terms of the merger of two
black holes.  At no time does the conformal geometry used as data
exhibit more than one hole. However, the horizon obtained by
integrating the single Einstein equation does have two holes at early
affine times, and just one hole at late affine times.  The authors
attribute these interesting features to the richness of the Einstein
equations; still, a good deal of foresight on their part must have
helped bring them to light.

{\bf References:}

[1] 
R.~A.~Matzner, H.~E.~Seidel, S.~L.~Shapiro, L.~Smarr, 
W.-M.~Suen, S.~A. Teukolski and J. Winicour, Geometry of a Black Hole 
Collision, {\it Science} {\bf 270},  pp 941-947 (1995).

[2]
Please check the {\it Science\/} article for references to several authors
that contributed computational results collected in the article. 

[3]
L. Lehner, N.~T. Bishop, R. G\'{o}mez, B. Szil\'{a}gyi and J. Winicour, Exact 
Solutions for the Intrinsic Geometry of Black Hole Collisions, gr-qc/9809034

\vfill
\pagebreak

\section*{\centerline {LIGO project update}}
\addtocontents{toc}{\protect\smallskip}
\addcontentsline{toc}{subsubsection}{\it  LIGO project update, by 
David Shoemaker}
\begin{center}
    David Shoemaker, MIT\\
\htmladdnormallink{dhs@ligo.mit.edu}
{mailto:dhs@ligo.mit.edu}
\end{center}
\parindent=0pt
\parskip=5pt

LIGO Installation is once again the focus of our efforts in the LIGO
Lab. Because the activities are so hands-on, this Update is mostly in
graphical form with the text serving as captions for the photographs
which can be found at
\htmladdnormallink{ftp://ligo.ligo.caltech.edu/pub/mog/figures29jan98.pdf}
{ftp://ligo.ligo.caltech.edu/pub/mog/figures29jan98.pdf}

We start (Fig. 1) with a view from space of the Livingston
Observatory. The 4km arms are visible (as the clearing of the forest),
and in the detailed view the 'X' shaped main building can be made
out. Descending from angel to airplane height, we now see clearly the
high-bay space containing the interferometer components on the right,
the covered beam tubes in the background and to the right, and the
entrance and offices in the right foreground. (The 'overpass' and the
black water tank are fire precautions.)

A view inside the high bay (Fig 3), this time from Hanford, shows a
number of the test-mass vacuum equipment chambers (the taller vertical
cylinders), some 'HAM' multipurpose chambers (the two to the right),
the main beam tubes (the large horizontal tubes to the far left and
the right background), and a few of the many electronics racks.
Navigating around the equipment involves lots of walking and climbing
of stairs!

Several 2km sections of the beam tube (Fig 4) have been successfully
'baked out'---heated to drive off excess water and other contaminants.
We see here a section of beam tube, wrapped in insulation, a heavy
cable snaking across the floor to deliver current for heating. The
concrete cover is arch-shaped.

A very significant effort is now underway in both Hanford and
Livingston to install the seismic isolation system. Vacuum cleanliness
requires 'bunny suits' (Fig 5); all of the equipment placed in the
vacuum must be cleaned and baked as well, to guard against
contamination of the mirrors. Fig 6 is a view after installation, with
the bottom table, cylindrical masses and somewhat hidden springs
between them, the top 'optics' table, and counterweights (emulating
the final load) all visible.

The test masses, fused silica 25cm in diameter, are carefully
characterized in a metrology interferometer (Fig 7), and mounted in a
cage with a simple wire loop. A detail of the point of departure of
the wire is seen at the bottom left. The optical losses are determined
by the polish and coating (and its cleanliness), and the mechanical
losses are the point of connection with thermal noise, and so
excruciating attention must be given to every detail.

The optics are installed (Fig 8) in the vacuum system and given an
initial alignment sufficiently precise that the reflected beam will be
correctly aligned to within one beam tube radius (0.5 m) over the
length of the beam tube (4 km). Fancy surveying!

Our last image (Fig 9) shows a part of the optical table carrying the
Pre-Stabilized Laser and some of the Input Optics (the University of
Florida's contribution). The cylindrical vacuum system contains the
frequency reference cavity for the laser, and the rectangular block of
fused silica (developed at Stanford) is an optical cavity used in
transmission to 'clean' the optical beam.

Our schedule calls for first tests of an interferometer using just the
optics in the main building for this summer, with the full 4km paths
included in the fall of '99. Please visit one of the sites if you are
in the vicinity; contact and other information can be found at
\htmladdnormallink{http://www.ligo.caltech.edu}{http://www.ligo.caltech.edu}.

\vfill
\eject

\section*{\centerline {
``Bicentenary of the Cavendish Experiment" Conference}}
\addtocontents{toc}{\protect\smallskip}
\addcontentsline{toc}{subsubsection}{\it  
Bicentenary of the Cavendish Experiment, by Riley Newman}
\begin{center}
    Riley Newman, University of California Irvine\\
\htmladdnormallink{rdnewman@uci.edu}
{mailto:rdnewman@uci.edu}
\end{center}
\parindent=0pt
\parskip=5pt

The measurement of G has become a popular occupation following the
announcement a few years ago of   widely discrepant new G values --- most
notably a value reported by the German PTB (NIST-equivalent) lab which is
more than half a percent away from the accepted "CODATA" value (see volume 4
of this Newsletter).   Thus the celebration of the 200th anniversary of
Cavendish's G measurement with a conference in London last November,
sponsored by IOP and organized by Terry Quinn of the BIPM and Clive Speake,
drew a lively crowd of G measurers to Cavendish's turf.  Twelve groups
actively pursuing G measurements were represented; six of these announced
new or updated G values:
\\
\\
\begin{tabular}{|l|l|c|c|c|} \hline

Lab  &  G $ \times 10^{11}$   &  $ (ppm)$  &  $ G-G_{CODATA}\ (\sigma) $
&  $ G - G_{PTB} \  (\sigma) $ \\ \hline \hline
New Zealand MSL  &  6.6742(6)   &  90   &  1.5   &  -51   \\ \hline
Zurich  &  6.6749(14)  &  210  &  1.4  &  -27  \\ \hline
Wuppertal  &  6.6735(9)(13)  &  240  &  0.5  &  -25  \\ \hline
JILA  &  6.6873(94)  &  1400  &  1.6  &  -3  \\ \hline
BIPM  &  6.683(11)  &  1650  &  0.9   &  -3  \\ \hline
Karagioz (Russia)  &  6.6729(5)   &  75   &  0.3   &   -57  \\ \hline
\hline
Luther/Towler 1982  &  6.6726(5)  &  64  & -- &   -58 \\ \hline
PTB 1995 &  6.71540(56)  &  83   & 42 & --  \\ \hline
\end{tabular}

Here the comparisons with CODATA and PTB in standard deviations reflect
uncertainties in both numbers compared.  The last two lines give the 1982
result of Luther and Towler (on which the CODATA value is based, after
doubling the assigned uncertainty), and the puzzling 1995 PTB result.

All six new results are higher than but roughly consistent with the CODATA
value.   The PTB value remains a mystery, not to be lightly dismissed ---
the CODATA committee will have difficult decisions to make in its next round
of assessments!  No lab yet feels it has surpassed the 64 ppm accuracy  that
Luther and Towler assigned to their 1982 measurement, although a number of
groups target accuracy of 10 ppm or better.

The twelve approaches to G measurement are remarkable in their variety -- no
two are very similar in technique.  Heedful of Kuroda's caution about the
perils of anelasticity, all but two of the experiments either avoid the use
of a torsion fiber or use a fiber in a mode such that its internal strains
are negligible.
The New Zealand MSL lab compares the torque on a torsion pendulum with that
due to an electrostatic force which is in turn calibrated in terms of the
angular acceleration it produces on the pendulum in a separate experiment.
The Zurich group uses a beam balance to weigh kilogram masses in the
presence of mercury filled steel tanks -- this group anticipates greatly
increased accuracy when some systematics issues are resolved.
Wuppertal measures the effect of source masses on the spacing of a pair of
suspended masses which form a microwave Fabry-Perot cavity, and aims for
accuracy better than 100 ppm.
JILA uses its free-fall gravimeter to measure the change in g produced by a
movable tungsten ring mass.
BIPM uses a torsion balance suspended by a thin flat metal strip; its
dominant torsional restoring force is gravitational, thus minimizing
anelastic dangers.  BIPM plans to use its instrument in two ways to measure
G, in both a static displacement mode and a "time of swing" dynamic mode,
aiming for a solidly reliable measurement at a 100 ppm level.
The Russian lab uses a torsion pendulum in the classic dynamic mode used by
Luther and Towler; it has been troubled in the past by poorly understood
drifts.

Work in progress was reported by additional groups:  Luther at LANL is
developing an instrument which will use a bifilar suspension which, like the
BIPM strip suspension, has a restoring torque which is dominantly
gravitational in origin thus circumventing anelasticity issues.  The
University of Washington and Irvine labs are both building instruments which
use thin plate pendulums suspended in nearly pure quadrupole gravitational
field gradients produced by special source mass configurations.  The
Washington approach elegantly avoids fiber-related problems by servoing its
pendulum to a continuously rotating platform whose measured periodic angular
acceleration  reflects that of the pendulum, while the pendulum fiber never
twists significantly.  The Irvine instrument uses the classic "time of
swing" dynamic method, operating at 2K with a high Q fiber whose anelastic
effects should be sufficiently small and well understood to not limit the
measurement's accuracy.  Both the JILA group and the Taiwan group of W.T. Ni
discussed plans for G measurements using a scheme like Wuppertal's but using
an optical rather than microwave interferometer as distance gauge.  The SEE
project was presented, which hopes to determine G and test other aspects of
Newtonian gravity using measurements of ``horseshoe" trajectories of test
masses projected toward a field mass within a long cylindrical space
capsule.  Development at the Politecnico di Torino of a G measurement using
a pendulum swinging between two mass spheres was discussed.

The conference also featured fascinating talks by G. Gillies and I. Falconer
on the history of G measurements and the painfully shy but highly skilled
man Cavendish.   Thibault Damour lectured on the theoretical importance of
measurements of G and its possible dependencies on mass composition,
distance, and time.  Thibault reminded us that, contrary to popular belief,
G is NOT the least well known fundamental constant -- that distinction
belongs to the strong coupling constant!

{\bf G measurement lab contacts:}  

New Zealand: 
\htmladdnormallink
{t.armstrong@irl.cri.nz}
{mailto:t.armstrong@irl.cri.nz}

Zurich: 
\htmladdnormallink
{schlammi@physik.unizh.ch}
{mailto:schlammi@physik.unizh.ch}

Wuppertal:
\htmladdnormallink
{meyer@wpos7.physik.uni-wuppertal.de}  
{mailto:meyer@wpos7.physik.uni-wuppertal.de},

JILA: 
\htmladdnormallink
{Fallerj@jila.colorado.edu}
{mailto:Fallerj@jila.colorado.edu}

BIPM:  
\htmladdnormallink
{tquinn@si.bipm.fr}
{mailto:tquinn@si.bipm.fr}

Karagioz (Russia): 
\htmladdnormallink
{irtrib@cityline.ru}{mailto:irtrib@cityline.ru}

Ni (Taiwan): 
\htmladdnormallink
{wtni@phys.nthu.edu.tw}{mailto:wtni@phys.nthu.edu.tw}

SEE: 
\htmladdnormallink
{ASanders@utk.edu}{mailto:ASanders@utk.edu}

Politecnico di Torino: 
\htmladdnormallink
{demarchi@polito.it}{mailto:demarchi@polito.it}

Irvine:  
\htmladdnormallink
{rdnewman@uci.edu}{mailto:rdnewman@uci.edu}

Washington:
\htmladdnormallink
{gundlach@dan.npl.washington.edu}
{mailto:gundlach@dan.npl.washington.edu}

\vfill
\pagebreak

\section*{\centerline {Eighth Midwest gravity meeting}}
\addtocontents{toc}{\protect\smallskip}
\addtocontents{toc}{\bf Conference reports:}
\addtocontents{toc}{\protect\smallskip}
\addcontentsline{toc}{subsubsection}{\it  
Eighth Midwest gravity meeting, by Richard Hammond}
\begin{center}
    Richard Hammond, North Dakota\\
\htmladdnormallink{rhammond@plains.NoDak.edu}
{mailto:rhammond@plains.NoDak.edu}
\end{center}
\parindent=0pt
\parskip=5pt

The Eighth Midwest Relativity Meeting was held September 24---25
at North Dakota State University in Fargo. The transparencies are
 electronically published on the MWRM8 website

\htmladdnormallink
{http://www.phys.ndsu.nodak.edu/mrm8.html}
{http://www.phys.ndsu.nodak.edu/mrm8.html} 

The meeting was kicked
off with a bang by Beverly Berger's characterization of
singularities for generic matter. In her work she assumed the
existence of only one Killing vector, and discussed the outlook
for the case of no symmetries. Bob Wald presented an intriguing
discussion of the Path Integral in quantum gravity, and after
emphasizing our incomplete understanding of both the wave function
and the path integral, compared it with the parameterized
Sch\"odinger quantum mechanics, in the context of tunneling, and
showed there was a critical parameter for tunneling. Rich Hammond
demonstrated how a cosmological term gradient breaks the principle
of equivalence and gave an upper bound for its gradient based on
laboratory results.

Jean Krisch discussed cosmology in D+1 dimensions, and showed that
as D increased the Planck temperature decreased. Jim Wheeler
explained his conformal theory of gravity in 4+4 dimensions, while
Terry Bradfield used a ``compensating'' field to obtain conformal
invariance. Mike Martin examined gauge fields in the context of a
classical unified theory of gravity and electromagnetism.

Gabor Kunstatter made his debut at MWRMs with a discussion on the
entropy of black holes (why is it so large?) in gravitation with a
scalar field. Brett Taylor showed how a scalar field can make the
black hole temperature zero, and Leopoldo Zayas discussed
procedures for obtaining black hole entropy using string theory
methods.

Bill Hiscock gave an overview of the OMEGA project, which, if
chosen, will be one of NASA's new smaller missions which will
orbit an interferometer around earth-moon. He emphasized that
there are six known white dwarf binaries that OMEGA should detect.
Shane Larson calculated the noise in the interferometer without
resorting to the usual long wavelength approximation. Ted Quinn
calculated the force on a scalar particle in curved space
including the radiation reaction force. Ken Olum explained fast
travel in terms of negative mass, or Casimir-type energy. We all
know the picture with Einstein standing in front of the blackboard
with ``$R_{i\kappa}=0?''$. Dwight Vincent gave a fascinating account
of this picture, from its photo-shoot in 1931 at the Mt. Wilson
Observatory, to modern commercialization of it. Discussions
focused on the meaning of the question mark, and what physics was
actually being questioned.

Charro Gruver derived the material action for gravitation with
torsion, and showed how to obtain the correct conservation law for
angular momentum plus spin. Bill Pezzaglia derived equations of
motion for particles with spin in the presence of torsion by
generalizing the Lagrangian to include an area term.

Robert Mann presented an interesting exact solution for the N-body
problem in two dimensional gravity with a scalar field, and
showed, for example, that when the Hamiltonian becomes large with
respect to $3mc^2$ the relativistic effects are fully evident.
Thomas Baumgarte defined a new conformal 3-metric to modify the
ADM formalism, and discussed geodesic slicing vs. harmonic slicing
with regard to  numerical solutions.

Ed Glass gave an illuminating discussion of the Vaiyda metric, and
showed that a generalization leads to both a null fluid and
string fluid. Ivan Booth discussed boundary terms  and Steve
Harris discussed the future causal boundary and multiply warped
spacetimes. Mike Ashley discussed the properties of the a-boundary
as a topological object. Homer Ellis explained space-time-time,
and `dark hole' solutions with singularities. Ian Redmount
examined quantum field theory states in Robertson Walker cosmology
and claimed that particles can only be well defined in an open
universe. This created a lively discussion with Wald pointing out
that his results are probably not valid for massless particles
(such as photons). The discussion continued when Berger, Mann, and
John Friedmann joined the fray.

Marc Paleth reported on the Wigner function and the relation
between its peaks and regions of high correlation. Another lively
discussion followed when Zayas questioned the $\hbar\rightarrow0$
limit. Andreas Zoupas explained how environmental de-coherence
arises from the master equation and the reduced equation. James
Geddes gave a detailed account of the measure in the path integral
from the point of view of a collection of subsets.

Excitement grew as meeting neared its end with John Friedmann's
fascinating discussion of  how, in a window between $10^9$ and
$10^{10}$, perturbations in a rotating neutron star can grow.
Finally, Abraham Ungar used the law of Einstein velocity addition
to generalize the laws of motion, and suggested tests for his
theory.

\vfill
\pagebreak

\section*{\centerline {GWDAW '98}}
\addtocontents{toc}{\protect\smallskip}
\addcontentsline{toc}{subsubsection}{\it  
GWDAW '98, by Sam Finn}
\begin{center}
    Lee Samuel Finn, Penn State,
\htmladdnormallink{lsf@gravity.phys.psu.edu}
{mailto:lsf@gravity.phys.psu.edu}
\end{center}
\parindent=0pt
\parskip=3pt

On 19--21 November 1998, nearly 100 researchers from Australia, Asia,
Europe and North America gathered in Central Pennsylvania to attend
the Third Gravitational Wave Data Analysis Workshop,
 held under the auspices of Penn State's Center for Gravitational
Physics and Geometry.

Like the first GWDAW, hosted by MIT in December 1996, GWDAW'98 was
organized as a workshop. There were only a small number of short
(typically 15 minutes) invited talks, each of which introduced or
commented on an outstanding challenge in gravitational wave data
analysis. Each talk was followed by extended, moderated discussion
(typically 45 minutes) among all the participants. Each invited
speaker was admonished by the organizing committee to speak to the
future: to look, not backward to the achievements of yesterday, but
forward to the challenges ahead and how they might be addressed.

To provide focus for the meeting, the organizing committee identified
four different areas of study that would benefit from intense, focused
discussion in a workshop setting. Speakers were chosen and the
workshop was organized around the subjects of data diagnostics, upper
limits and confidence intervals, LISA data analysis challenges, and
collaborative data analysis.

The workshop's first morning began an orientation: a series of status
reports given by representatives of the four different interferometer
projects currently under construction --- GEO~600 (B. S.
Sathyaprakash), LIGO (A. Lazzarini), TAMA~300 (N. Kanda and T.
Tanaka), and VIRGO (A. Vicere') --- with an emphasis on their
developing plans for data handling and analysis.

Following lunch, the participants turned to the discussion of {\em
  data diagnostics:\/} the use of the data channel itself as a
diagnostic monitor of both the instrument's health and the usefulness
of the signal channel for analysis. Dr.~S.~Vitale (University of
Trento) described a new, automated data quality monitoring system that
has been developed and installed on the AURIGA cryogenic acoustic
detector. Two components of this system were of particular interest to
the workshop participants: the first was the requirements it placed on
the distribution of the signal channel output over hour-long
intervals; the second was the requirement of consistency in the
excitation amplitude of the antenna's two resonant modes.

Following this discussion, Dr. S. Mukherjee (Penn State) described the
development of a {\em Kalman Filter\/} that can extract from the
signal channel of an interferometric detector the amplitude and phase
of the mirror suspension violin modes. Passing gravitational waves do
not excite these modes, while technical mechanical noise sources that
move the interferometer test masses do; thus, the suspension modes are
a sensitive monitor of mechanical disturbances masquerading as
gravitational waves. The wide-ranging discussion that followed
included the use of the Kalman predictor as a means of reducing the
data dynamic range and, possibly, the removal of the lines before
analysis for gravitational wave signals.

The afternoon was rounded out by a presentation by Drs. N.~Kanda
(Miyagi University of Education) and D.~Tatsumi (Institute for Cosmic
Ray Research, University of Tokyo), who outlined of the TAMA detector
calibration scheme and discussed some of the the effects of
calibration error on the ability to reliably detect and identify the
parameters of gravitational wave sources.

Despite their unprecedented sensitivity and bandwidth, there is no
guarantee that the first generation of interferometric detectors will
detect directly any sources. Nevertheless these instruments can --- at
the very least --- set interesting and provocative upper limits on
source strengths and rates. The second morning of the workshop was
given-over to a discussion of the use of data from gravitational wave
detectors to set upper limits.

Four talks punctuated a very animated discussion. The morning began
with a tutorial, including several worked examples, on the
construction of confidence intervals and credible sets, given by Dr.
S. Finn (Penn State University). Two lessons were evident in this
presentation: the first, that the analysis involved in the
construction of upper limits and confidence intervals is as involved
as that which goes into deciding upon detection or measuring parameter
values, and the second, that the procedures for constructing
confidence intervals and credible sets involve choices, and that these
affect quantitatively the upper limits or confidence intervals
derived.

Following this tutorial, Dr. S. Mohanty (Penn State University)
described how, even in the absence of a known waveform, one can set
upper-limits on the gravitational wave strength from gamma-ray bursts
or other potential gravitational wave sources associated with an
astrophysical trigger.  After a lengthy discussion period, Dr. B.
Allen (University of Wisconsin, Milwaukee) discussed an analysis of
the LIGO 40M prototype November 1994 data set for binary inspiral
signals. Focusing on the largest ``detected'' signal in the data set,
this collaboration placed what might be termed an upper limit on the
possible sets of upper limits that could arise from a more detailed
analysis. Finally, Dr. M. Papa (Albert Einstein Institute) inaugurated
a discussion of several different methods for searching for weak
periodic gravitational wave signals in long stretches of data.

The afternoon of the first day focused on data analysis challenges
associated with LISA: the proposed Laser Interferometer Space Antenna.
LISA is currently being considered by NASA as a joint ESA/NASA
mission, with a potential launch data as early as 2007. LISA has a
much greater immediate potential for detecting directly gravitational
waves from known astrophysical sources; however, the challenges of
LISA data analysis are different than for ground-based interferometers
and it is important to show now that LISA's promise can be realized
through the analysis of the data it would collect. Four talks, spread
over an afternoon of discussion, focused the workshop's attention on
LISA's current status (Dr. S. Vitale, University of Trento), data
analysis challenges (Dr. R. Stebbins, JILA and University of
Colorado), ability to direct the attention of astronomers to imminent
activity in different parts of the sky (Dr. A. Vecchio, Albert
Einstein Institute), and what turns out to be the unimportance of
gravitational lensing of cosmological sources for LISA (Dr. C. Cutler,
Albert Einstein Institute).

The final session of the conference was devoted to an animated
discussion on collaborative data analysis.  There is a general
agreement that the joint analysis of data from all simultaneously
operating detectors will lead to more constraining upper limits,
greater confidence in reported detections, greater measurement
precision, and more information about observed sources.  There are,
however, practical impediments to collaborative data analysis.  These
include the different ``personalities'' of the experimental apparatus,
deriving from their great complexity When different teams design,
build, commission and characterize their instruments, where resides
the common knowledge required to carry out an analysis that makes
sophisticated use of the multiple data streams?

This problem has been tackled by the acoustic detector community and
some of the lessons they have learned were discussed by Dr. G.
Pizzella (University of Rome "Tor Vergata"). Other communities face
the identical problem: Dr. B. Barish (Caltech) described the mechanism
that the neutrino detection community has developed for data sharing.
Finally, Dr. R.~Weiss (MIT) closed out the workshop with a final
presentation on the organization of data analysis in the COBE project
and his own perspective on the importance of cooperative data
analysis. Weiss identified what may be one of the more difficult
problems we face: the cultural transition from the model of the
scientist as an individual entrepreneur, who keeps hold of an idea and
the credit for it, to the collaborative model, where ideas are shared,
the community accepts the credit for the accomplishments, and the
participants take their reward from being part of the community.

\vfill
\pagebreak
             
\section*{\centerline {Bad Honnef seminar}}
\addtocontents{toc}{\protect\smallskip}
\addcontentsline{toc}{subsubsection}{\it  
Bad Honnef seminar, by Alan Rendall}
\begin{center}
    Alan Rendall, Albert Einstein Institute\\
\htmladdnormallink{rendall@aei-potsdam.mpg.de}
{mailto:rendall@aei-potsdam.mpg.de}
\end{center}
\parindent=0pt
\parskip=5pt

A seminar with the title \lq Mathematical Problems in General Relativity\rq\
took place in Bad Honnef, Germany, from 7th to 11th September 1998. This
seminar, with about sixty participants from eighteen countries, was
organized by Herbert Pfister (University of T\"ubingen) and Bernd Schmidt (AEI
Potsdam) and financially supported by the Heraeus Foundation. It was
primarily aimed at graduate students and so strong emphasis was put on the
pedagogical nature of the talks. At the same time it was intended to give a
point of entry to recent results in the area of mathematical relativity.

A mathematical understanding of general relativity requires knowledge of
the solution theory of the Einstein constraints and evolution equations
and the corresponding mathematical background. Robert Beig gave an
introduction to the constraints together with a discussion of identifying
spacetime Killing vectors in terms of initial data and giving a
four-dimensional characterization of special solutions of the constraints,
such as the multiple black hole solutions of interest in numerical
relativity. Oscar Reula treated the basic theory of the evolution equations.
He also presented a more general account of the nature of hyperbolicity of
systems of partial differential equations and new results on
writing the Einstein equations expressed in Ashtekar variables in symmetric
hyperbolic form.

A subject which was given particular attention was that of asymptotically flat
vacuum spacetimes with Helmut Friedrich talking for four hours on his
program to investigate the consistency of the classical conformal picture
with the field equations and Alan Rendall talking for four hours on the
theorem of Christodoulou and Klainerman on the nonlinear stability of
Minkowski space. Friedrich presented his results
on the stability of de Sitter and anti-de Sitter spacetimes as well as
new developments concerning restrictions on asymptotically flat initial
data related to smoothness of null infinity. As explained in talks by
Gabriel Nagy, insights from the anti-de Sitter case were important in the
recent existence theorem for the initial boundary value problem for the
vacuum Einstein equations by him and Friedrich. Rendall explained some of
the analytical techniques used in the Christodoulou-Klainerman proof
such as energy estimates (also prominent in Reula's talk), the Bel-Robinson
tensor, bootstrap arguments and the null condition. Lars Andersson showed
how some of these techniques, in particular the Bel-Robinson tensor, have
been applied to a class of cosmological spacetimes in his work with Vincent
Moncrief on the stability of the Milne model. This opens up the possibility
that the Christodoulou-Klainerman result may not stand in splendid
isolation much longer. Yvonne Choquet-Bruhat, in
a talk on geometrical optics expansions for the Einstein equations, told
how this reveals an \lq almost linear\rq\ property of these equations,
exceptional among hyperbolic systems, which is related to the null
condition.

Matter was not neglected at the conference either. Herbert Pfister and
Urs Schaudt described their progress towards constructing solutions of
the Einstein-Euler equations with given equation of state representing
rotating stars. Lee Lindblom talked on the inverse problem of
reconstructing the equation of state given data on masses and radii of
corresponding fluid bodies. In constructing fluid bodies it is always
wise to keep an an eye on the corresponding Newtonian problem. J\"urgen
Ehlers gave an introduction to his mathematical formulation of the
Newtonian limit which can be used to give a conceptually clear approach
to this. Gerhard Rein summarized our present knowledge on the gravitational
collapse of collisionless matter, including his recent numerical work with
Rendall and Jack Schaeffer on the boundary between dispersion and black hole
formation for this matter model. {}From the point of view of exact solutions,
Gernot Neugebauer spoke on the inverse scattering method and Dietrich Kramer
described an approach to producing null dust solutions.

The fact that the whole seminar took place in the Physics Centre of the
German Physical Society in Bad Honnef and that all participants were
accomodated in that building provided ample opportunity for formal
and (particularly on the evening with free beverages) less formal
discussions.

\end{document}